# Spectral GUI^R for Automated Tissue and Lesion Segmentation of T1 Weighted Breast MR Images

*Prajval Koul**

## Abstract:

We present Spectral GUI^R, a multiplatform breast MR image analysis tool designed to facilitate the segmentation of fibro glandular tissues and lesions in T1 weighted breast MR images via a graphical user interface (GUI). Spectral GUI^R uses spectrum loft method [1] for breast MR image segmentation. Not only is it interactive, but robust and expeditious at the same time. Being devoid of any machine learning algorithm, it shows exceptionally high execution speed with minimal overheads. The accuracy of the results has been simultaneously measured using performance metrics and expert entailment. The validity and applicability of the tool are discussed in the paper along with a crisp contrast with traditional machine learning principles, establishing the unequivocal foundation of it as a competent tool in the field of image analysis.

## Keywords:

## 1. Introduction:

Nowadays, Magnetic Resonance images are used into a variety of applications ranging from rudimentary analyses to extensive research. Due to the workload involved in manual analyses of medical images like MR images, an interactive image analysis interface is a highly sought-after demand. The current prevalent trend among people handling and working with this data is to apply some form of machine learning or its derivative for analysis [2]. Although, machine learning is an innovative tool to use, it comes with its own flaws. For instance, supervised learning brings along with it the infamous scenario of 'Big Data Deluge' which can't be resolved effectively because the so called 'data' here is extremely indispensable to supervised learning. Although the amount of data is lesser in semi-supervised or reinforced learning, for that matter, but this carry along data dents deeply the portability of the system it serves. This flaw is recovered using unsupervised learning up to some extent, but here also data is beheld closely to the system to define certain parameters. These flaws don't seem to be a big deal due to the existence of cheap hardware storage units, but this comes to surface once the system is to be transported elsewhere. This gives us the motivation to create a real time MR image analysis tool which is easy to port and light on the running machine at the same time. The tool has been carefully crafted using MATLAB's Graphical User Interface Developing Environment (GUIDE) to make it more lucid, interactive and aesthetic for the end user. The following sections explain in detail the principles followed, methods used, and the designing procedure followed while making of the Spectral GUI^R.

*Department of Computer Science and Engineering, National Institute of Technology Delhi, Delhi (110040), India.

# 2. Theory and Experiment:

Every breast MR image intensity histogram has the following features:
1. Adipose tissue intensity spectrum.
2. Fibro glandular tissue intensity spectrum.

Both above properties appear as peaks in the overall intensity distribution. Rudimentary understanding of the above statement suggests that there must exist a loft in this spectrum which, in this case, is in the form of a valley. Acknowledgement of this fact is the motivation behind the concept of spectrum loft method. Some trivial assumptions are made (although not compulsory in nature), to efficiently locate the intensity threshold for segmentation, as follows:

### 2.1. *Assumptions:*
1. Intensity of adipose tissues can't fall below an absolute value of 300 (for a uint16 image format).
2. Similarly, the intensity of fibro glandular tissues can't exceed an absolute value of 800 (for a uint16 image format).

These assumptions are based on an extended survey of breast MR images of over 100 subjects and can be easily scaled up or down according to the image format. Hence, the valley of the histogram should lie well within these two intensity boundaries.

Note: No assumptions are required in case of lesion segmentation as opposed to tissue segmentation.

### 2.2. MR image pre-processing:

#### *2.2.1. Removal of Ghost Artefacts*
Ghost artefacts are the unwanted low intensity signals present in MR image exterior. If not removed, they hinder proper segmentation of image. Ghost artefacts are removed by the following sequence of operations. The MR image is binarized using a low intensity threshold. This binarized image is eroded using a structuring element in order to remove the loosely connected components in the MR image exterior. The eroded image is finally dilated using the same structuring element to get an image mask devoid of ghost artefacts. This mask is used get the original image back, but with a clear background.

#### *2.2.2. Removal of Speckle Noises*
Speckle noises are the high intensity low frequency stray signals present in the MR image interior. These noises generally give a granular appearance to the image, interfering with its further processing. Minimization of speckle noises is carried out by using Anisotropic Diffusion [3]. Not only does it preserve edges, it also helps avoid blurring. The equations used in this method are mentioned below:

$$\frac{\partial}{\partial t} v(y,t) = div(cc(y,t).\nabla v(y,t)) \quad ....(1)$$

Where, cc(y, t) is the gradient or diffusion constant which is given as [II],

$$cc(y,t) = \exp\left[-\left(\frac{y}{k}\right)^2\right] \quad ...(2)$$

Where, k is the gradient magnitude threshold parameter. Previously, Perona et. al. [II] have defined parameter k as user specific which is taken by hit and trial method and the value of k is different for each image. This approach has been slightly modified to make it useful for all images [4]. In this modification, the whole image has been taken as window to calculate the mean and standard deviation and the edge magnitude parameter k has been redefined with a modified mathematical expression. The mathematical formula of the modified edge magnitude parameter can be expressed in a generalized form as,

$$k = 2\log(m \times n) \frac{\sqrt{mean}}{std} \quad ...(3)$$

This method works for all types of images. Speckle index is calculated to assure and verify the proper removal of these noises.

### *2.2.3. Correction of Bias Field (Intensity Inhomogeneity)*

Correction of intensity inhomogeneity is done by using level set approach [5-6]. Here, the inhomogeneity in the image intensities is characterized as an innate property of the image itself, a biasing field or simply a bias field. The image is modelled as:

$$F = gT + N \qquad \ldots(4)$$

Where *'T'* is true image, *'g'* is the component that accounts for the gain field or intensity inhomogeneity and *'n'* is the additive noise. The true image *'T'* measures an intrinsic physical property of the objects being imaged, which is, therefore, assumed to be piecewise constant. The gain field *'g'* is assumed to be a smooth slowly varying function. The varying noise 'n' can be assumed as zero mean Gaussian noise [5]. The bias field corrected image is computed as the quotient *'F/ g'*. In the end, the image is smoothened in order to get an image with homogeneous intensities, which completes the pre-processing.

After completion of pre-processing, segmentation is carried out on the pre-processed image. As we all know, segmentation lays the premise for analyses like anatomical study of breast, breast density estimation and quantitative assessment and breast diseases' monitor. Here, the novel generic approach *'Spectrum Loft Method'* for segmenting out the fibro glandular tissues as well as lesions from breast MR images has been described, analyzed and transformed into an interactive tool which can be used by users belonging to any level of the technical hierarchy.

# 3. Methodology:

The making of the Spectral GUI$^R$ is very easy as it requires only the code to be transformed into an interface-based system rather than the entire previous history it had initially worked upon. Before discussing about the intricacies of the GUI makeover, the underlying method used, which is the spectrum loft method [1], is discussed briefly. This method, in itself, is complete and self-dependent in nature as it doesn't require any sort of human intervention at any intermediate step of its execution.

## 3.1. Spectrum Loft Method:

After pre-processing the required image, an MR image intensity histogram is generated. This is the key step involved. The following algorithm is applied in order to achieve desired results.

**Algorithm**:

1. Select the empirical intensity threshold boundaries.
2. Find all the local minima between the above chosen boundaries.
3. Choose the optimum threshold from the above values having minimum frequency.

Once the optimum threshold has been found, it can be used as an intensity threshold for segmenting out the MR images.

For segmenting out the lesions, a rather tactical modification of the basic spectrum loft analysis has been implemented. The optimum threshold is chosen from the histogram of a complete DCE breast MR image which is then used to segment out lesions sequentially from the chest segmented image. Not only does this method guarantee higher computational efficiency, but also helps get accurate results. This is a beautiful amendment to the basic premise of spectrum loft method which outgrows its formerly described domain [1].

All the segmentation results are, thereby, compared to the ground truth, which was crafted by an expert radiologist, using standard performance measures like DSC, JI and DI, which are computed and compared [1], along with expert technical entailment in case of lesions.

## 3.2 Making of the GUI:

The GUI not only covers every aspect of the underlying method used, it also embeds within itself the prerequisite pre-processing required for the images to be worked upon. Addition of this feature to it makes it fully automated in every respect.

The method, along with all its nuances, has been converted into an interactive GUI with the help of MATLAB's GUDE. GUIDE was used to elicit the basic skeleton of the front end of the interface, which consists of elements like axes, buttons, panels etc. The skeleton consists of 2 panels, 3 axes, 4 buttons and a pair of radio buttons. Each element of the front end has its own call back function, set to null by default. These call back functions were modified to obtain the desired interface.

Following is the sequence of steps involved in the working of the GUI:

1. Using the 'Select Image' button, browse an image from the directory to be segmented and select it. It should be noted that tissue segmentation requires pre-contrast breast MR images where as lesion segmentation requires DCE or post-contrast images.
2. Select the type of segmentation one wishes to perform, according to the type of image selected in the previous step from the 'Segmentation Type' radio button panel.
3. Click the 'Segment' button in the window to fetch the results.

The above sequence helps generate the segmented image along with a complementary histogram of the selected image so that the user can 'visualise' the spectral loft and hence, appreciate the overall method.

# 4. Results and Discussion:

The study has been done on bilateral axial Tl-weighted MR images of over 160 subjects of both segmentation types. The subjects have been imaged in prone position covering the entire chest (breasts with pectoral muscles) using Siemens Magnetom Avanto (TIM 76 x 18), version Syngo MK 1317, a 1.5 T MR scanner. The dataset has been acquired from Rajiv Gandhi Cancer Institute and Research Centre, Delhi, India. It consists of 160 axial slices per subject. Each bilateral axial slice consists of 448 x 448 voxels in 16-bit greyscale format. The slices are spaced 1mm apart.

## 4.1 Segmentation Revisited:

After successful pre-processing of the selected image, spectrum loft method comes into play and helps segment out the image as desired.

To establish the accuracy and elegance of the method, following segmentation results are shown along with respective frequency spectra.

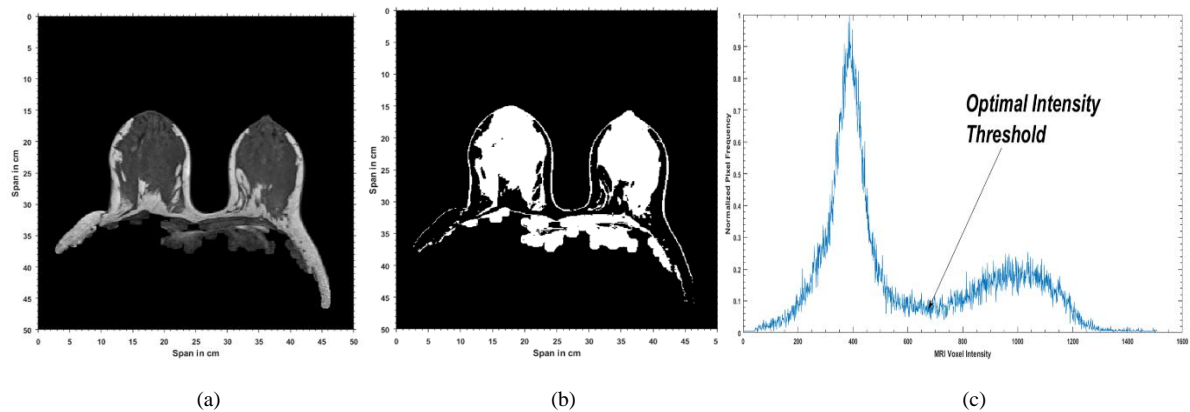

(a)      (b)      (c)

Fig. 1.(a) A randomly selected Class D Pre-contrast breast MR image. (b) Fibro glandular Tissue Segmentation using Spectrum Loft Method (c) Intensity Spectrum of the MR image.

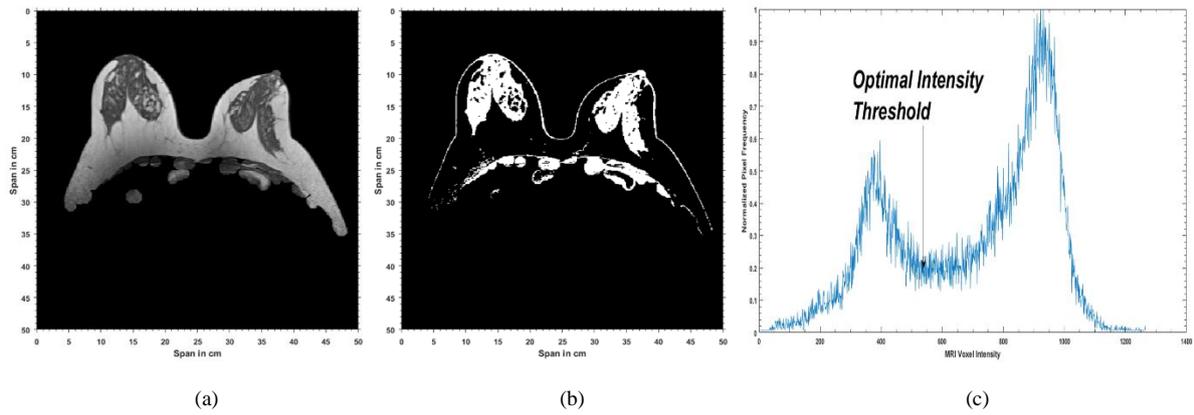

(a)          (b)          (c)

Fig. 2. (a) A randomly selected Class C Pre-contrast breast MR image. (b) Fibro glandular Tissue Segmentation using Spectrum Loft Method (c) Intensity Spectrum of the MR image.

In the above figures, the first image is an un-pre-processed pre contrast MR image. The second image is the final segmented image using the above stated method. The third image is the frequency spectrum of the pixel intensities of the initial image.

Similar convincing results for lesion segmentation which were obtained, are as follows:

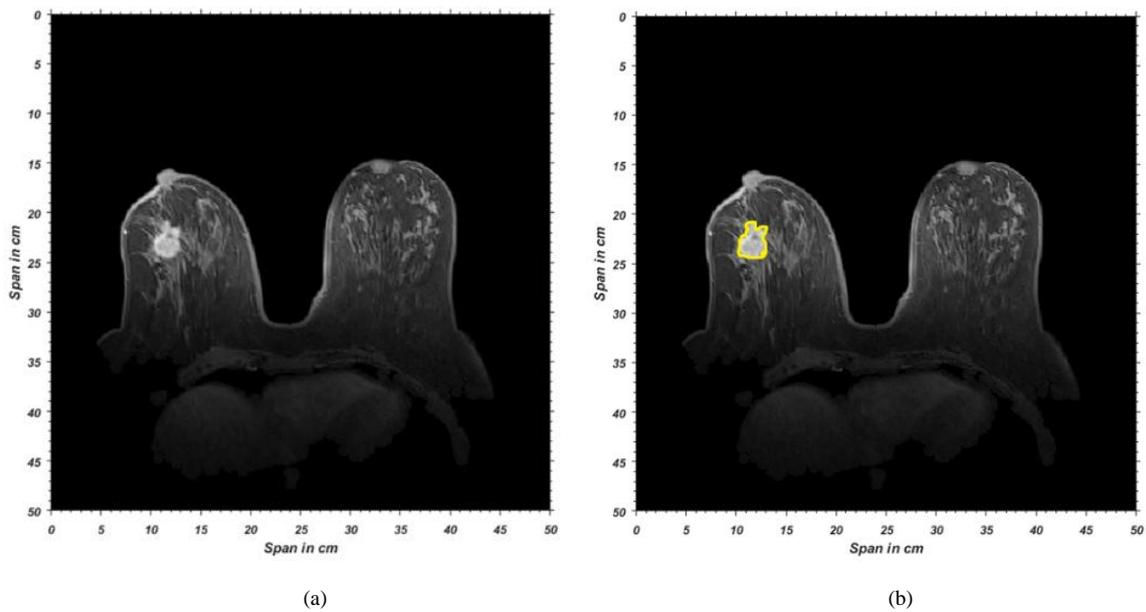

(a)          (b)

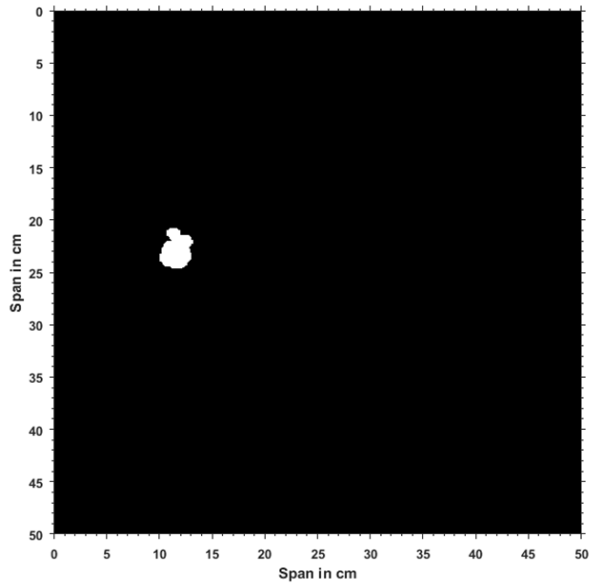
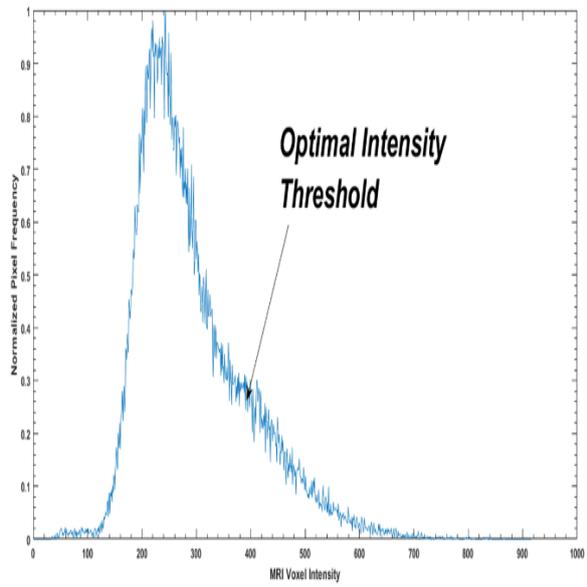

(c)                      (d)

Fig. 3. (a) A randomly chosen DCE breast MR image with a visible lesion (b) Lesion contoured (yellow) by expert entailment (for reference) (c) Segmented Lesion by Spectrum Loft Method (d) Intensity Spectrum of MR Image

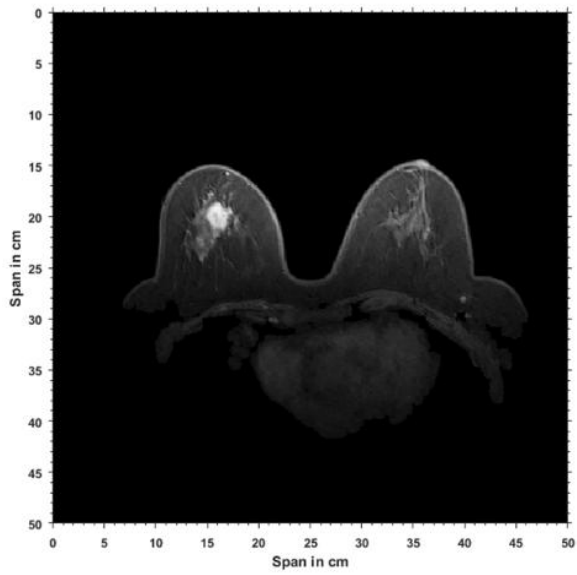
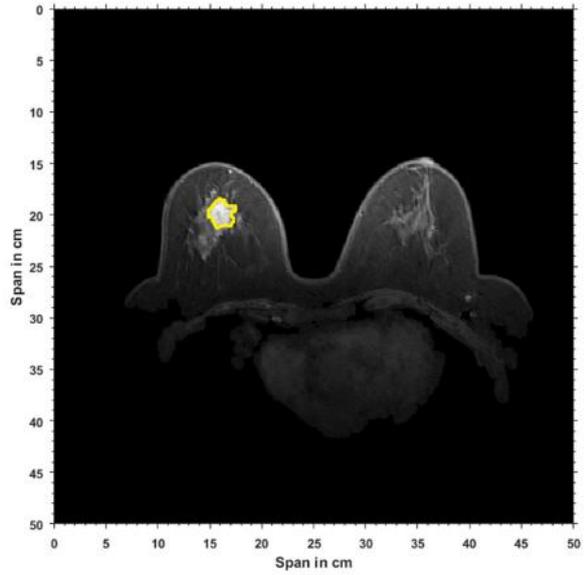

(a)                      (b)

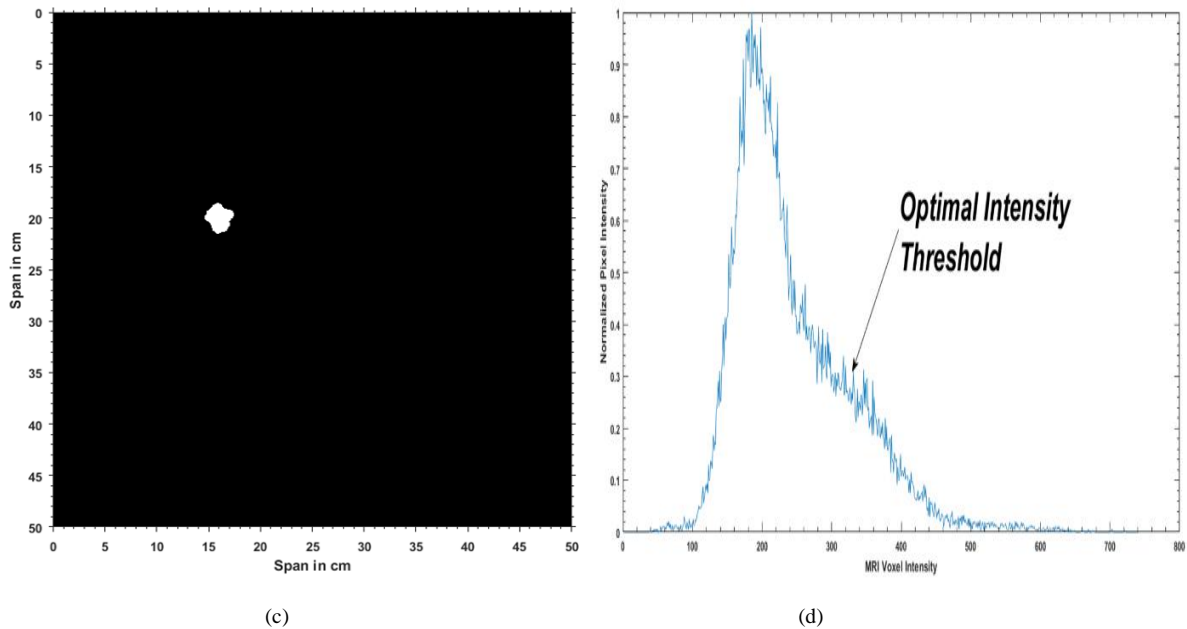

| (c) | (d) |

Fig. 4. (a) A randomly chosen DCE breast MR image with a visible lesion (b) Lesion contoured (yellow) by expert entailment (for reference) (c) Segmented Lesion by Spectrum Loft Method (d) Intensity Spectrum of MR Image

In the above figures, the first one is a post contrast MR image of the entire chest. This is followed by the second image which consists of a contour by an expert radiologist acting as a ground truth. The third image is the final segmented image using the above stated method. The fourth image is the frequency spectrum of the initial MR image.

Although, the valley in the intensity histogram of DCE MR images is not as pronounced as that of the histogram, this 'subtle' dip is enough for the method to be detect as well as segment out the lesions sequentially. The prominence of the dip doesn't depend upon how it appears in the frequency spectrum, but on the fact that how slyly it is obtained from the collaborating spectra under analysis.

## 4.2 Graphic User Interface:

The above results were individual entities which were arranged as shown above. To combine all these results and assemble them at one place, the GUI comes into play. The GUI lucidly interacts with the user and generates results in almost real time meticulously arranged in a window.

### 4.2.1 Tissue Segmentation:

Initially, the GUI appears in front of the end user in the following form.

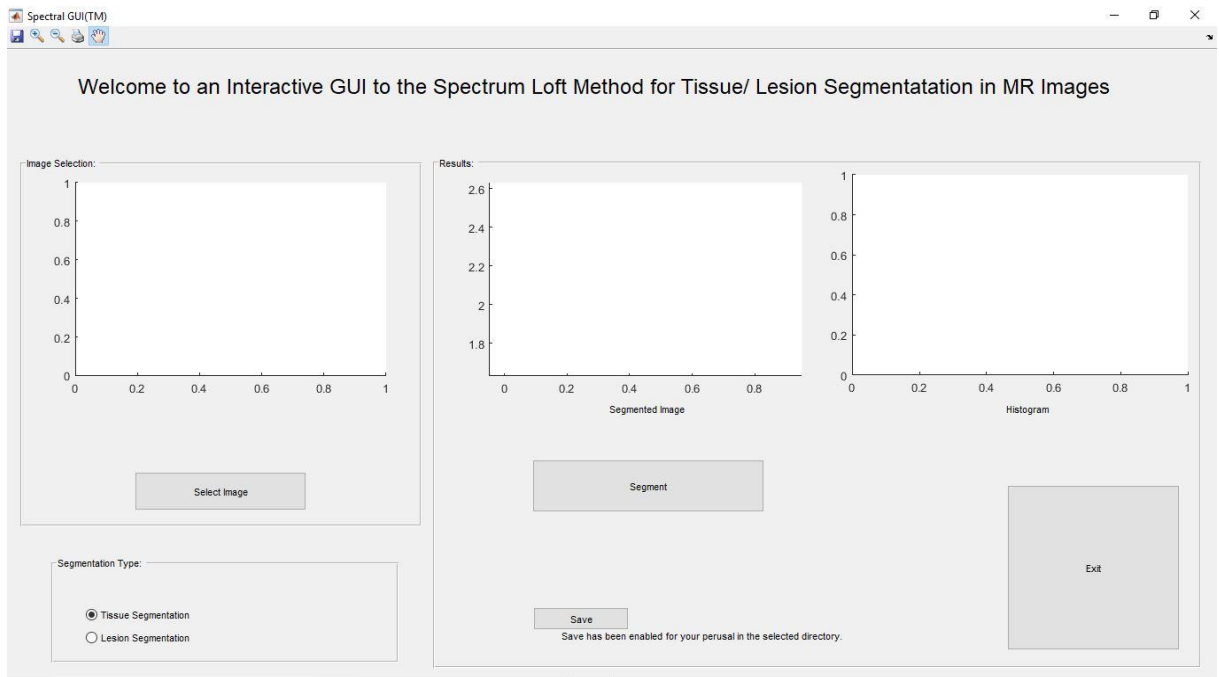

Fig.5.(a) Initial window that appears when the tool is launched.

On clicking the 'Select Image' button, a standard browse window appears where the respective image to be analyzed can be searched and selected to be loaded into the tool. Upon selection of the required image, the following changes take place in the initial state of the window.

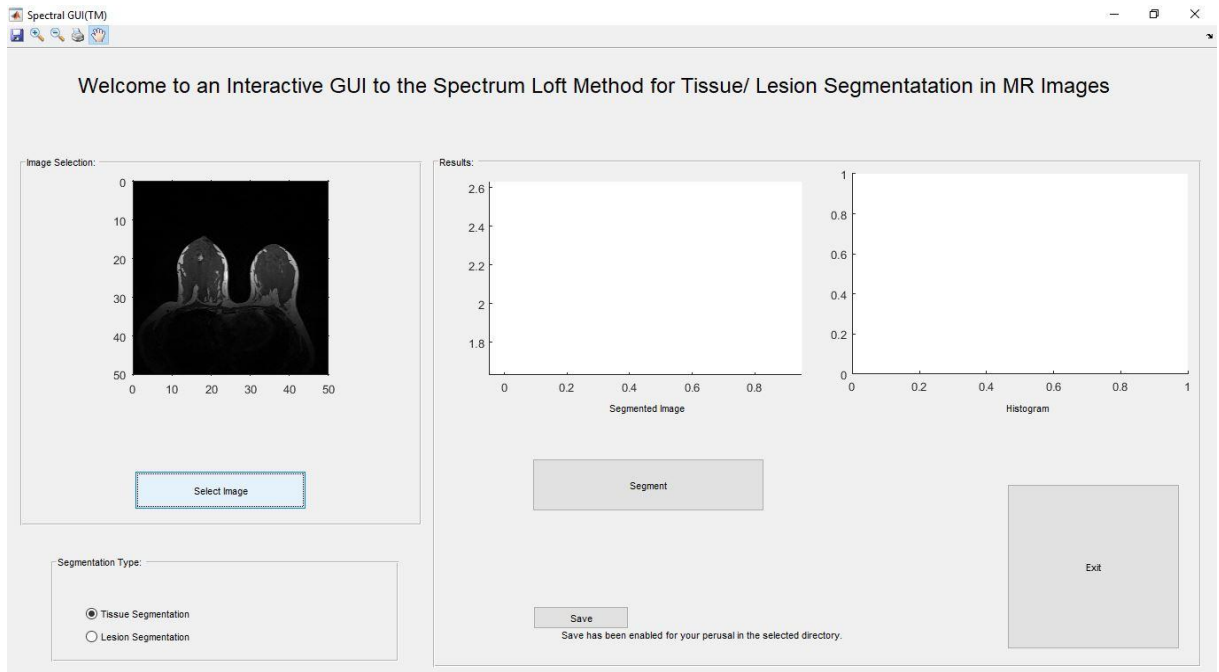

Fig. 5.(b) The image browsed in the previous step displayed in the first panel.

Now, the type of segmentation is selected from the 'Segmentation Type' radio button panel, as per the requirement of the image selected formerly. This means that if the image selected is pre-contrast MR image, 'Tissue Segmentation' is to be selected and if its DCE, 'Lesion Segmentation' is to be selected. Clearly, in this scenario, 'Tissue Segmentation' is to be selected. The selected choice will be highlighted in the way every standard radio button is, which can also be seen below.

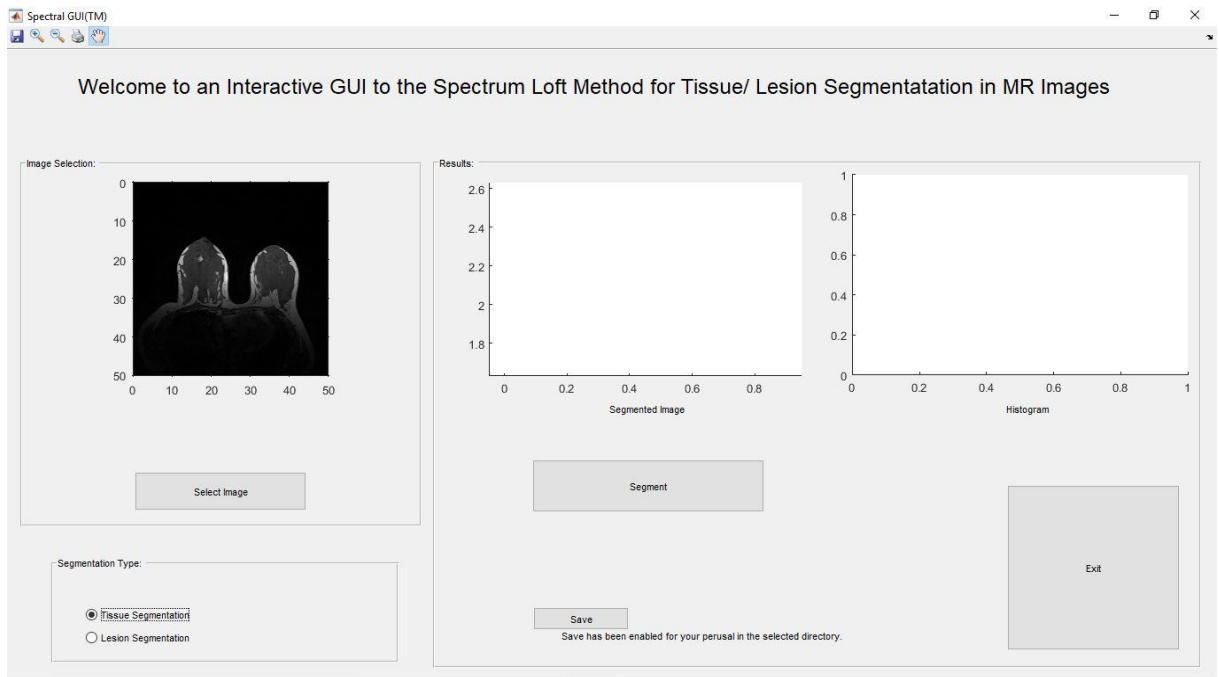

Fig. 5.(c) Segmentation type highlighted when clicked.

That's it. All prerequisites are taken care of. All one needs to do now is to click the 'Segment' button on the window and wait for a couple of seconds to get the desired result along with a complimentary intensity histogram. The following screenshot shows a sample output for an image selected initially.

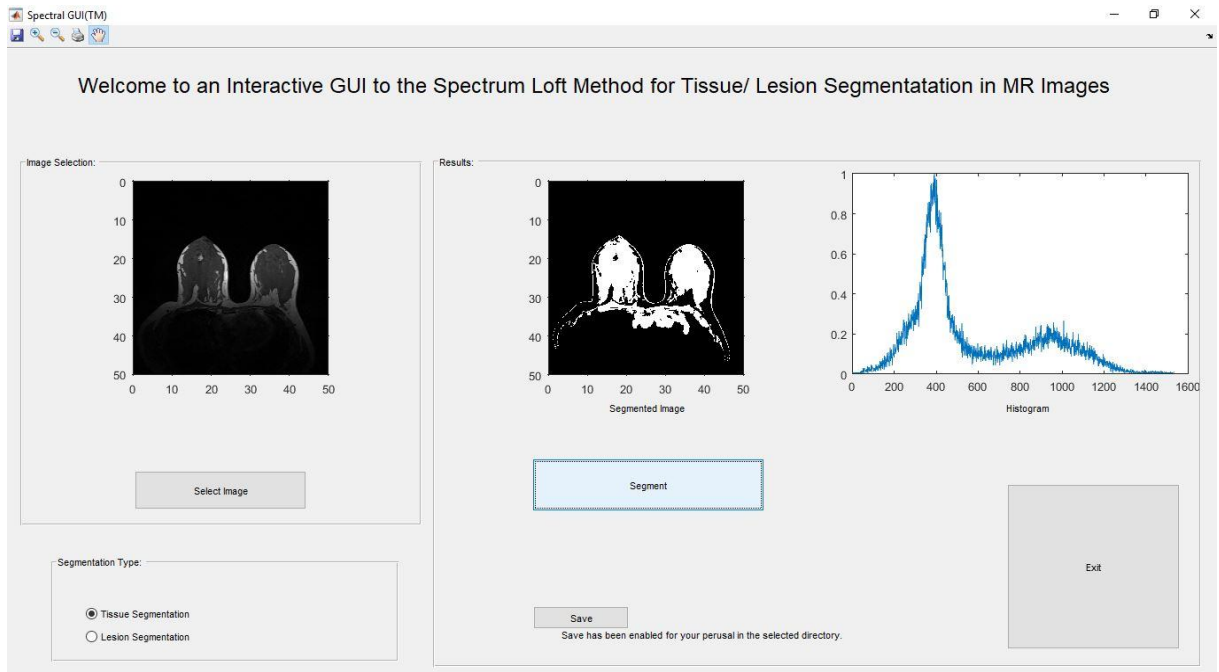

Fig. 5.(d) Fibro glandular Tissues segmented from the image along with intensity spectrum displayed in the respective panel when 'Segment' button is pressed.

Now, either the user can exit from the GUI using the 'Exit' button, or he can load another image using 'Select image' button and repeat the above cycle again. A complementary 'Save' feature has been included allowing the user to save the obtained results if desired.

Here, it should be noted that the modal time from clicking the 'Segment' button until getting the results is close to 3.40 seconds.

### 4.2.2 Lesion Segmentation:

Like the above case, the initial appearance of the GUI is as shown.

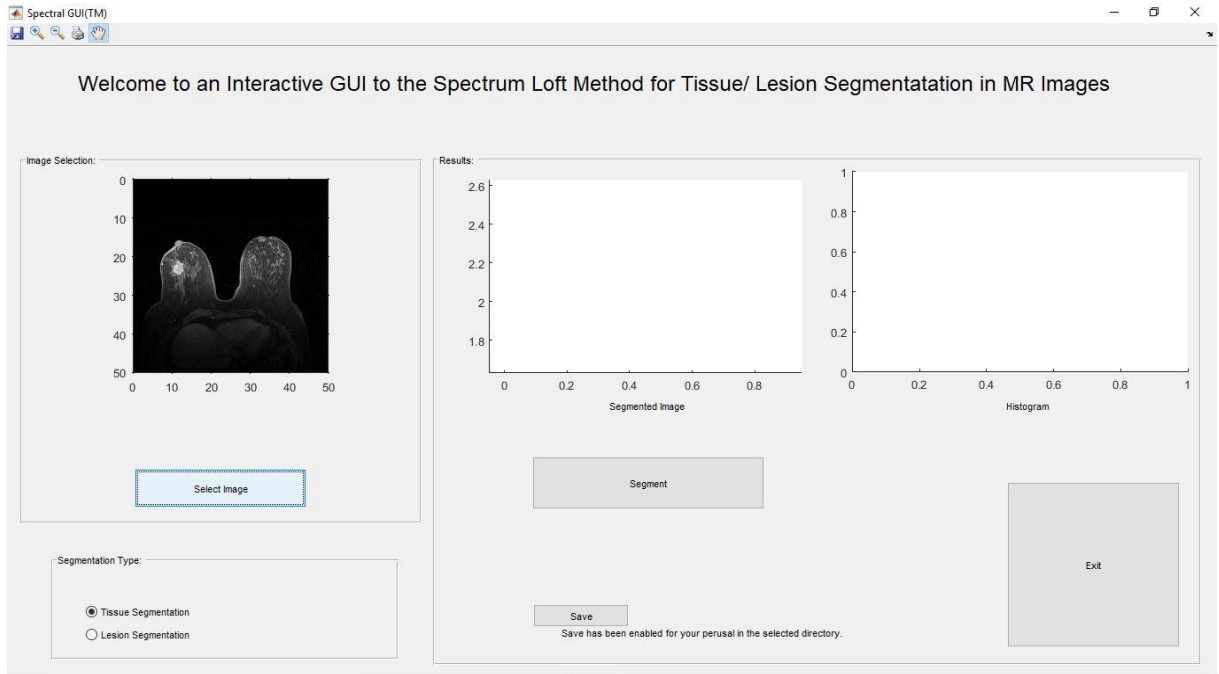

Fig. 6 (a) Initial window that appears when the tool is launched.

On clicking the 'Select Image' button, a standard browse window appears where the respective mage to be analyzed can be searched and selected to be loaded into the tool. Upon selection of the required image, the following changes take place in the initial state of the window.

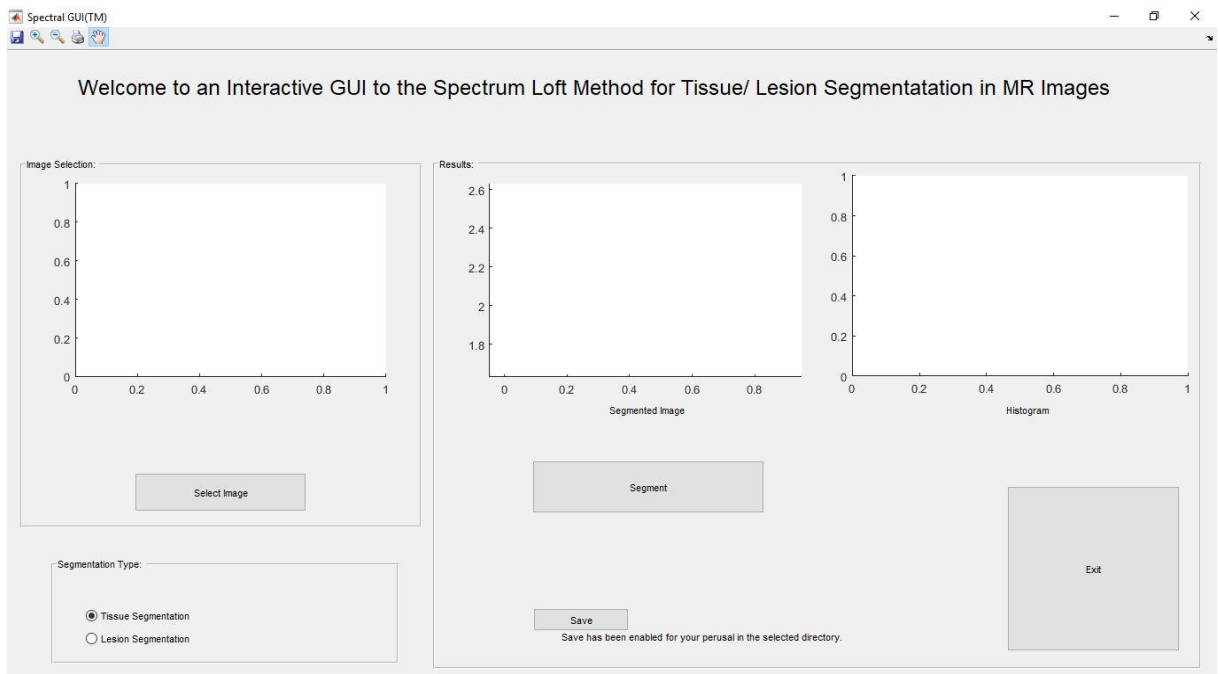

Fig. 6.(b) The image browsed in the previous step displayed in the first panel.

Now, the type of segmentation is selected from the 'Segmentation Type' radio button panel, as per the requirement of the image selected formerly. This means that if the image selected is pre-contrast MR image, 'Tissue Segmentation' is to be selected and if its DCE, 'Lesion Segmentation' is to be selected. Clearly, in this scenario, 'Lesion Segmentation' is to be selected. The selected choice will be highlighted in the way every standard radio button is, which can also be seen below.

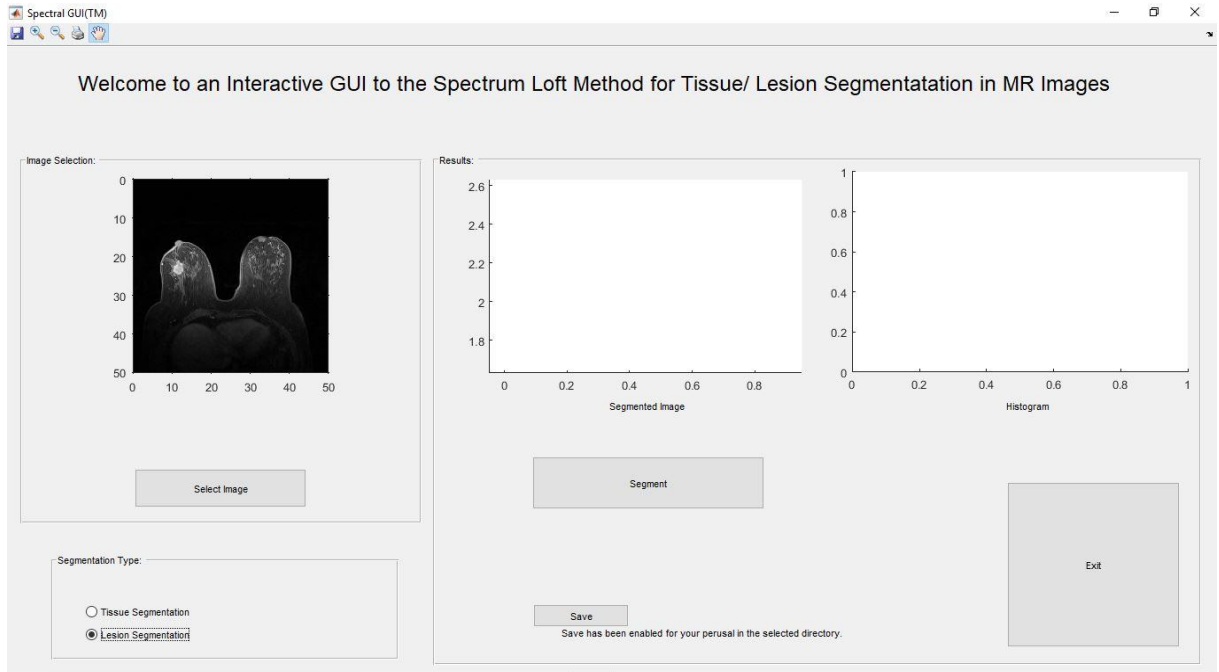

Fig. 6.(c) Segmentation type highlighted when clicked.

That's it. All prerequisites are taken care of. All one needs to do now is to click the 'Segment' button on the bottom right corner of the window and wait for a couple of seconds to get the desired result along with a complimentary intensity histogram. The following screenshot shows a sample output for an image selected previously.

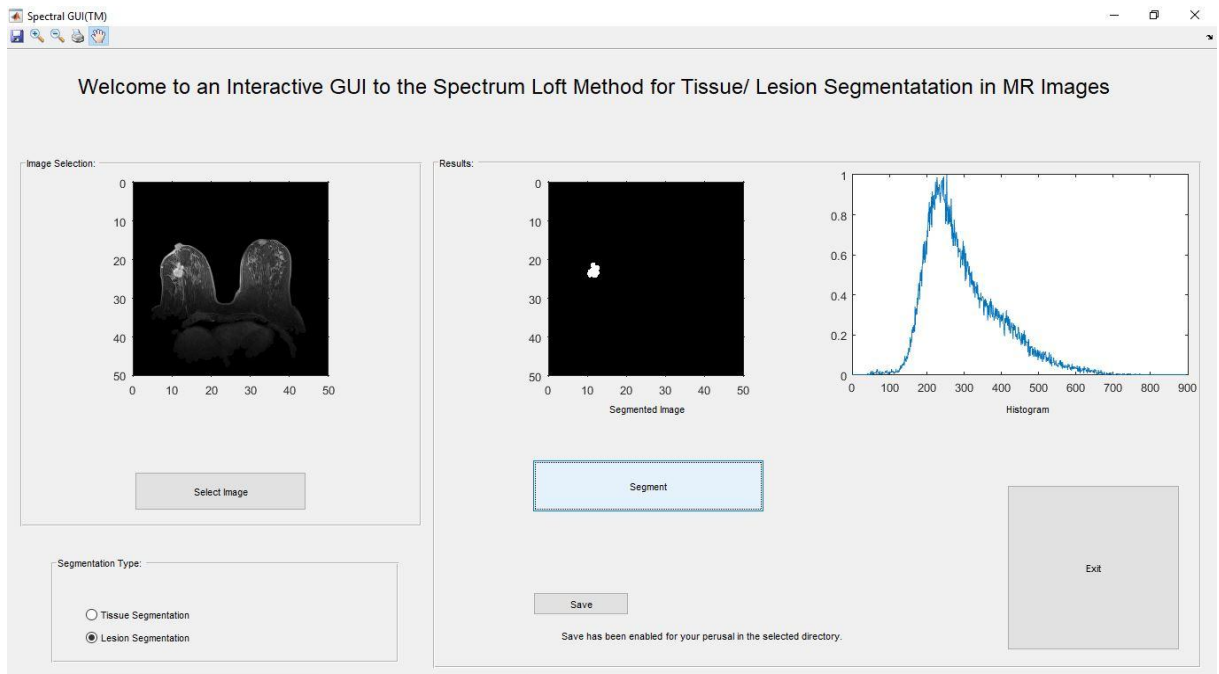

Fig. 6.(d) Lesion segmented out from the image along with intensity spectrum displayed in the respective panel when 'Segment' button is pressed.

Now, either the user can exit from the GUI using the 'Exit' button, or he can load another image using 'Select image' button and repeat the above cycle again. A complementary 'Save' feature has been included allowing the user to save the obtained results if desired.

Here, it should be noted that the modal time from clicking the 'Segment' button until getting the results is close to 3.59 seconds.

Following are some interesting points to be noted from the above discussion:

1. The method doesn't rely on machine learning in any form.
2. The method is inherently expeditious in nature.
3. The tool is fully automated.
4. The tool is easily portable in nature.
5. The tool is extremely lucid to use.
6. The tool fetches results in real time and hence, comes with negligible overheads.
7. The tool takes care of data pre-processing as well, which is a step former to the actual method used.

# 5. Conclusion:

It can be easily inferred from the above discussion that Spectral GIU$^R$ is a simple yet effective tool for breast MR image segmentation. Using a GUI based approach not only diversifies the applications of the method involved but also makes it an easy to use tool for any person irrespective of his technical background, e.g. doctors, technicians, students etc. It is extremely portable in nature because it doesn't require any backend database or a cohort of any kind. It stands apart from the crowd since it doesn't take the help of machine learning in any form whatsoever. Its enhanced domain induces a future scope of extending this method such that its rendered fruitful for more types of MR images as well as other imaging modalities.